\documentclass[11pt,a4paper,pdftex]{article}
\pdfoutput=1
\usepackage{jcappub}

\newcommand{\bi}{\bibitem}
\newcommand{\be}{\begin{eqnarray}}
\newcommand{\ee}{\end{eqnarray}}
\newcommand{\rar}{\rightarrow}

\title{Super-spinning compact objects generated by thick accretion disks}

\author{Zilong Li and Cosimo Bambi$^1$
\note{Corresponding author}}

\affiliation{Center for Field Theory and Particle Physics \& Department of Physics,\\
Fudan University, 220 Handan Road, 200433 Shanghai, China}

\emailAdd{zilongli@fudan.edu.cn}
\emailAdd{bambi@fudan.edu.cn}

\abstract{If astrophysical black hole candidates are the Kerr black holes predicted 
by General Relativity, the value of their spin parameter must be subject to the
theoretical bound~$|a_*| \le 1$. In this work, we consider the possibility that these
objects are either non-Kerr black holes in an alternative theory of gravity or exotic
compact objects in General Relativity. We study the accretion process when their 
accretion disk is geometrically thick with a simple version of the Polish doughnut 
model. The picture of the accretion process may be qualitatively different from the 
one around a Kerr black hole. The inner edge of the disk may not have the typical 
cusp on the equatorial plane any more, but there may be two cusps, respectively 
above and below the equatorial plane. We extend previous work on the evolution 
of the spin parameter and we estimate the maximum value of $a_*$ for the
super-massive black hole candidates in galactic nuclei. Since measurements of 
the mean radiative efficiency of AGNs require $\eta > 0.15$, we infer the 
``observational'' bound~$|a_*| \lesssim 1.3$, which seems to be quite independent 
of the exact nature of these objects. Such a bound is only slightly weaker 
than~$|a_*| \lesssim 1.2$ found in previous work for thin disks.}


\begin{document}

\maketitle

%%%%%%%%%%%%%%%%%%%%%%%%%%%%%%%

\section{Introduction}

In 4-dimensional General Relativity, an uncharged black hole (BH) is described by
the Kerr solution and it is completely specified by only two parameters: the mass
$M$ and the spin angular momentum $J$~\cite{nohair1,nohair2,nohair3}. A
fundamental limit for a Kerr BH is the bound $|a_*| \le 1$, where $a_* = J/M^2$ is
the BH spin parameter\footnote{Throughout the paper, we use units in which
$G_{\rm N} = c = 1$.}. For $|a_*| > 1$, there is no horizon, and the central singularity
is naked, which is forbidden by the weak cosmic censorship conjecture~\cite{wccc}.
Despite the apparent possibility of forming naked singularities from regular initial
data~\cite{dan}, any effort to overspin an existing Kerr BH to $|a_*| > 1$
seems to be doomed to failure~\cite{enrico1a,enrico1b}, and, even if created, a Kerr
naked singularity would be unstable and it should quickly decay~\cite{dotti,enrico2}.

At the observational level, there are at least two classes of astrophysical BH
candidates~\cite{nara}: stellar-mass objects in X-ray binary systems ($M \approx 5 -
20$~$M_\odot$) and super-massive BH candidates at the center of every normal
galaxy ($M \sim 10^5 - 10^9$~$M_\odot$). All these objects are thought to be
the Kerr BHs predicted by General Relativity, but their actual nature has still to be
verified~\cite{test,test2}. The Kerr paradigm requires $|a_*| \le 1$, but such a bound 
may be violated if BH candidates are either non-Kerr BHs in an alternative theory 
of gravity or exotic compact objects in General Relativity. Several groups have 
recently put forwards different proposals to test the Kerr-nature of astrophysical BH 
candidates~\cite{gw1,gw2,gw3,b0,jp1,naoki,jp2,b1,b2,cn1,b3,cn2,b4,kra,b5,b6,glg}. 
In general, there is a strong correlation between $a_*$ and possible deviations 
from the Kerr geometry, in the sense that observations may not be able to distinguish 
a Kerr BH from a non-Kerr object with different $a_*$.

In this context, it is interesting to get an estimate of the possible range of the
value of $a_*$. In Refs.~\cite{spin1a,spin1b}, it was shown that a thin accretion disk around
a non-Kerr compact object can spin the body up to $|a_*| > 1$. In Ref.~\cite{spin2},
it was presented an argument suggesting that the today spin parameters of the
super-massive BH candidates at the centers of galaxies is likely smaller than 1.2,
regardless of the exact nature of these objects. Future measurements of the
radiative efficiency of AGNs may put stronger constraints on both the spin
and the deformation parameter~\cite{spin3}. It has been also argued that
rapidly-rotating non-Kerr BHs may have topologically non-trivial event
horizons~\cite{topo0,topo1,topo2}.

In this work, we study the hydrodynamical structure of thick accretion disks in
non-Kerr backgrounds. We consider the simplest case of a marginally stable disk in the
framework of the Polish doughnut model~\cite{polish1,polish2}. We show that the
accretion process may be qualitatively different with respect to the Kerr case. The
accretion disk may have two cusps, one above and one below the equatorial plane, and
the gas of accretion may plunge from the cusps to the poles of the compact object. As
in the Kerr background, the accretion process from a thick disk can be potentially more
efficient than a thin disk to spin the central object up. We thus revise the result presented
in Ref.~\cite{spin2}, including the possibility that a super-massive BH candidate
has experienced a period of super-Eddington accretion and that its current spin
parameter still exceeds the equilibrium value for a thin accretion disk. The new bound
on the spin parameter of the super-massive BH candidates is $|a_*| \lesssim 1.3$,
only slightly weaker than the bound $|a_*| \lesssim 1.2$ found in~\cite{spin2}.

The content of the paper is as follows. In Section~\ref{s-pdm}, we review the
Polish doughnut model and we discuss the possible accretion scenarios in the
case of a Kerr background. In Section~\ref{s-nk}, we apply this model to describe
the hydrodynamical structure of accreting disks around non-Kerr BHs in a 
putative alternative theory of gravity and to exotic compact objects in General 
Relativity. We find the same qualitative picture and we believe this is the general 
picture for non-Kerr backgrounds. In Section~\ref{s-spin}, we study the evolution 
of the spin parameter resulting from the accretion process as a function of the 
deformation of the compact object. The discussion of our result and the current 
spin measurements is reported in Section~\ref{s-dd}.
If we assume a radiative efficiency $\eta > 0.15$,
which is a conservative lower bound on the mean radiative efficiency of AGNs
inferred by a few groups with the Soltan's argument, we find that the spin 
parameter of the super-massive BH candidates in galactic nuclei is 
$|a_*| \lesssim 1.3$. Summary and conclusions are reported in Section~\ref{s-c}.

\section{The Polish doughnut model \label{s-pdm}}

Geometrically thin and optically thick accretion disks around BHs are
described by the Novikov-Thorne model~\cite{nt}, which is the relativistic
generalization of the Shakura-Sunyaev model. Here, self-gravitation
of the disk and gas pressure are neglected, so the fluid elements follow
the geodesics of the background metric. The Novikov-Thorne model is
thought to work for moderate accretion rates, when the accretion luminosity
is between a few per cent to about 30\% the Eddington luminosity of the
object~\cite{mcc}. The Polish doughnut model was proposed in~\cite{polish1,polish2}
to describe non-self-gravitating disks when the gas pressure is
not negligible. Because of the gas pressure, the disk can be
geometrically thick and the fluid elements do not follow the geodesics of
the background metric.

The Polish doughnut model requires that the space-time is stationary and
axisymmetric. The line element can be written as
\be
ds^2 = g_{tt}dt^2 + g_{rr}dr^2 + g_{\theta\theta}d\theta^2
+ 2g_{t\phi}dtd\phi + g_{\phi\phi}d\phi^2 \, ,
\ee
where the metric elements are independent of the $t$ and $\phi$ coordinates.
The disk is modeled as a perfect fluid with purely azimuthal flow:
\be
T^{\mu\nu} = (\rho + P)u^\mu u^\nu + g^{\mu\nu} P \, , \quad
u^\mu = \left( u^t, 0 , 0 , u^\phi \right) \, ,
\ee
where $\rho$ and $P$ are, respectively, the energy density and the pressure.
The specific energy of the fluid element, $-u_t$, its angular velocity,
$\Omega = u^\phi/u^t$, and its angular momentum per unit energy,
$\lambda = -u_\phi/u_t$, are given by
\be
u_t = - \sqrt{\frac{g^2_{t\phi} - g_{tt}g_{\phi\phi}}{g_{\phi\phi} +
2\lambda g_{t\phi} + \lambda^2 g_{tt}}} \, , \quad
\Omega = - \frac{\lambda g_{tt} + g_{t\phi}}{\lambda g_{t\phi}
+ g_{\phi\phi}} \, , \quad
\lambda = - \frac{g_{t\phi} + \Omega g_{\phi\phi}}{g_{tt}
+ \Omega g_{t\phi}} \, .
\ee
Note that $\lambda$ is conserved for a stationary and axisymmetric
flow in a stationary and axisymmetric space-time~\cite{polish1}.
The disk's structure can be inferred from the Euler's equations, $\nabla_\nu
T^{\mu\nu} = 0$:
\be
a^\mu = - \frac{g^{\mu\nu} + u^\mu u^\nu}{\rho + P} \partial_\nu P \, ,
\ee
where $a^\mu = u^\nu \nabla_\nu u^\mu$ is the fluid's 4-acceleration. If the
pressure is independent of the $t$ and $\phi$ coordinates (which follows from
the stationarity and axisymmetry of the background) and if the equation of state
is barotropic ($\rho = \rho(P)$), $a^\mu$ can be written as a gradient of a
scalar potential $W(P)$:
\be
a_\mu = \partial_\mu W \, , \quad W(P)
= - \int^P \frac{dP'}{\rho(P') + P'} \, .
\ee
After some algebra, one can see it is possible to express $\Omega$ as a
function of $\lambda$, i.e. $\Omega = \Omega(\lambda)$, and integrate
the Euler's equation to get $W$\footnote{In the special case $\lambda = {\rm const.}$,
$\Omega$ is not constant, but Eq.~(\ref{eq-w}) is still correct and the integral
vanishes.}
\be\label{eq-w}
W = W_{\rm in} + \ln \frac{u_t}{u_t^{\rm in}}
+ \int_{\lambda_{\rm in}}^{\lambda}
\frac{\Omega d\lambda'}{1 - \Omega \lambda'} \, ,
\ee
where $W_{\rm in}$, $\lambda_{\rm in}$, and $u_t^{\rm in}$ are the potential,
the angular momentum per unit energy, and the energy per unit mass at the
inner edge of the fluid configuration. Here, $W_{\rm in}$, $\lambda_{\rm in}$, and
$u^t_{\rm in}$ can be replaced by the value of $W$, $\lambda$, and $u^t$ at
any other point of the fluid's boundary. In the Newtonian limit, $W$ reduces to
the total potential, i.e. the sum of the gravitational potential and of the centrifugal
one, and at infinity $W=0$.

If the background metric is known, there is only one unspecified function,
$\Omega = \Omega(\lambda)$, which characterizes the fluid's rotation.
In the zero-viscosity case, this function cannot be deduced from any equation,
and it must be given as an assumption of the model.
Imposing a specific relation between $\Omega$ and $\lambda$, we can
find the equipotential surfaces $W = {\rm const.} < 0$, i.e. the surfaces of
constant pressure, which represent the possible boundaries of the fluid
configuration. One of these surfaces may have one (or more) sharp cusp(s),
which may induce the accretion onto the compact object: like the cusp at
the $L_1$ Lagrange point in a close binary system, the accreting gas can
fill out the Roche lobe and then be transferred to the compact object. The
mechanism does not need the fluid's viscosity to work, so the latter may be,
at least in principle, very low.

A particularly simple case is the configuration with $\lambda = {\rm const.}$,
which is marginally stable with respect to axisymmetric perturbations
(the criterion for convective stability is simply that $\lambda$ does not have
to decrease outward). In this specific case, the integral in Eq.~(\ref{eq-w})
vanishes and
\be\label{eq-w2}
W = \ln (-u_t) + {\rm const.}
\ee
In the Kerr space-time, we find five qualitatively different scenarios~\cite{polish2}:
\begin{enumerate}
\item $\lambda < \lambda_{\rm ms}$, where $\lambda_{\rm ms}$ is the
angular momentum per unit energy of the marginally stable equatorial circular
orbit (or innermost stable circular orbit, ISCO). No disks are possible, as there
are no closed equipotential surfaces.
\item $\lambda = \lambda_{\rm ms}$. The local minimum of $W$ corresponding
to the disk's center is located on the equatorial plane at the marginally stable
radius. However, it is not really a minimum, but a flex. The disk exists as an
infinitesimally thin unstable ring.
\item $\lambda_{\rm ms} < \lambda < \lambda_{\rm mb}$, where
$\lambda_{\rm mb}$ is the angular momentum per unit energy of the
marginally bound equatorial circular orbit. There are many
stationary configurations without a cusp and one disk with a cusp on the equatorial
plane, located between the marginally bound and the marginally stable radius.
Accretion starts when the gas fills out all the equipotential surface with the cusp.
\item $\lambda = \lambda_{\rm mb}$. The cusp is located on the equatorial plane
and belongs to the marginally closed equipotential surface $W = 0$. Accretion is
possible in the limit of a disk of infinite size.
\item $\lambda > \lambda_{\rm mb}$. No accretion is possible, as there are
no equipotential surfaces $W \le 0$ with a cusp.
\end{enumerate}

\begin{figure}
\begin{center}
\includegraphics[type=pdf,ext=.pdf,read=.pdf,width=7.6cm]{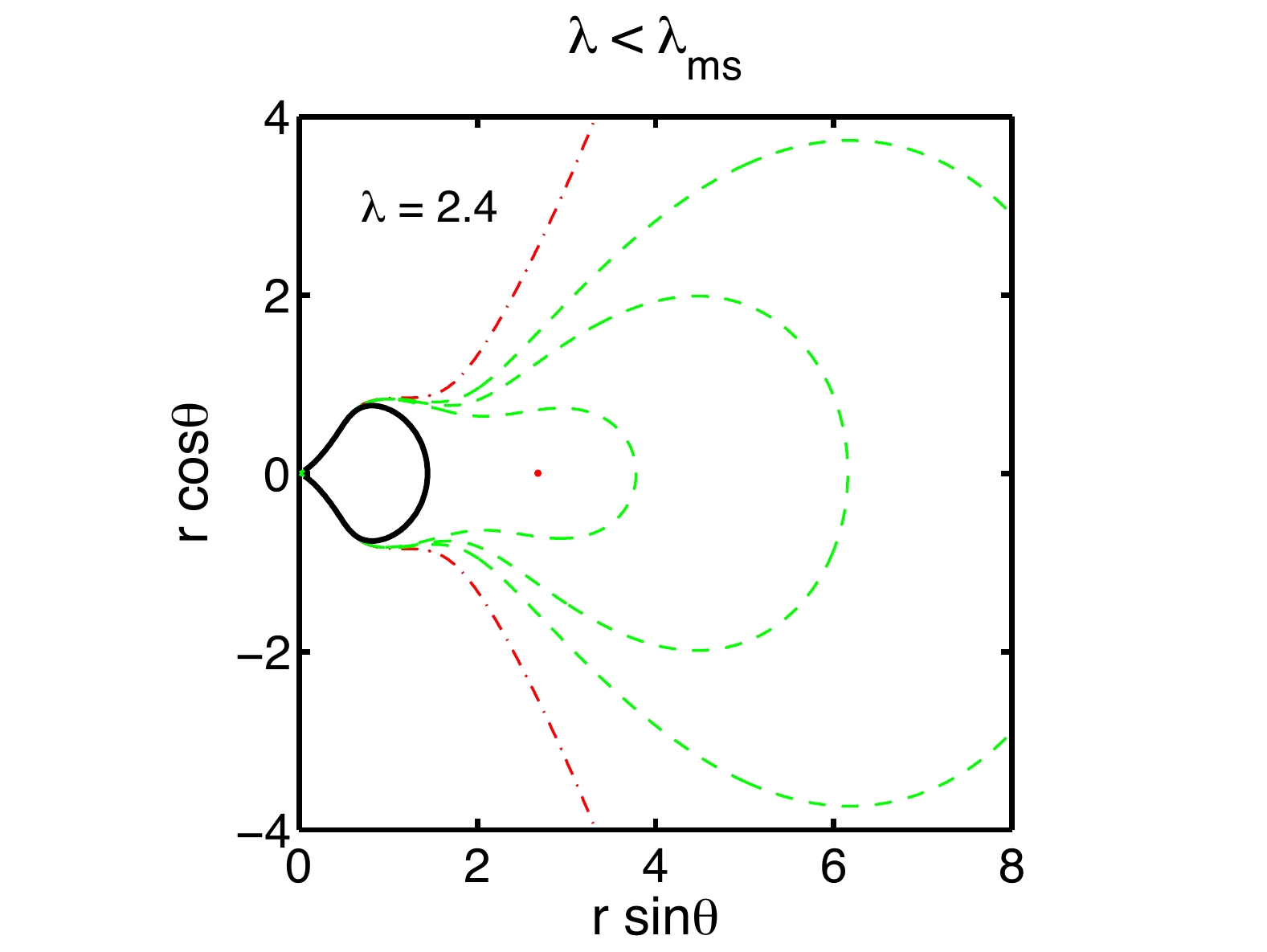}
\includegraphics[type=pdf,ext=.pdf,read=.pdf,width=7.6cm]{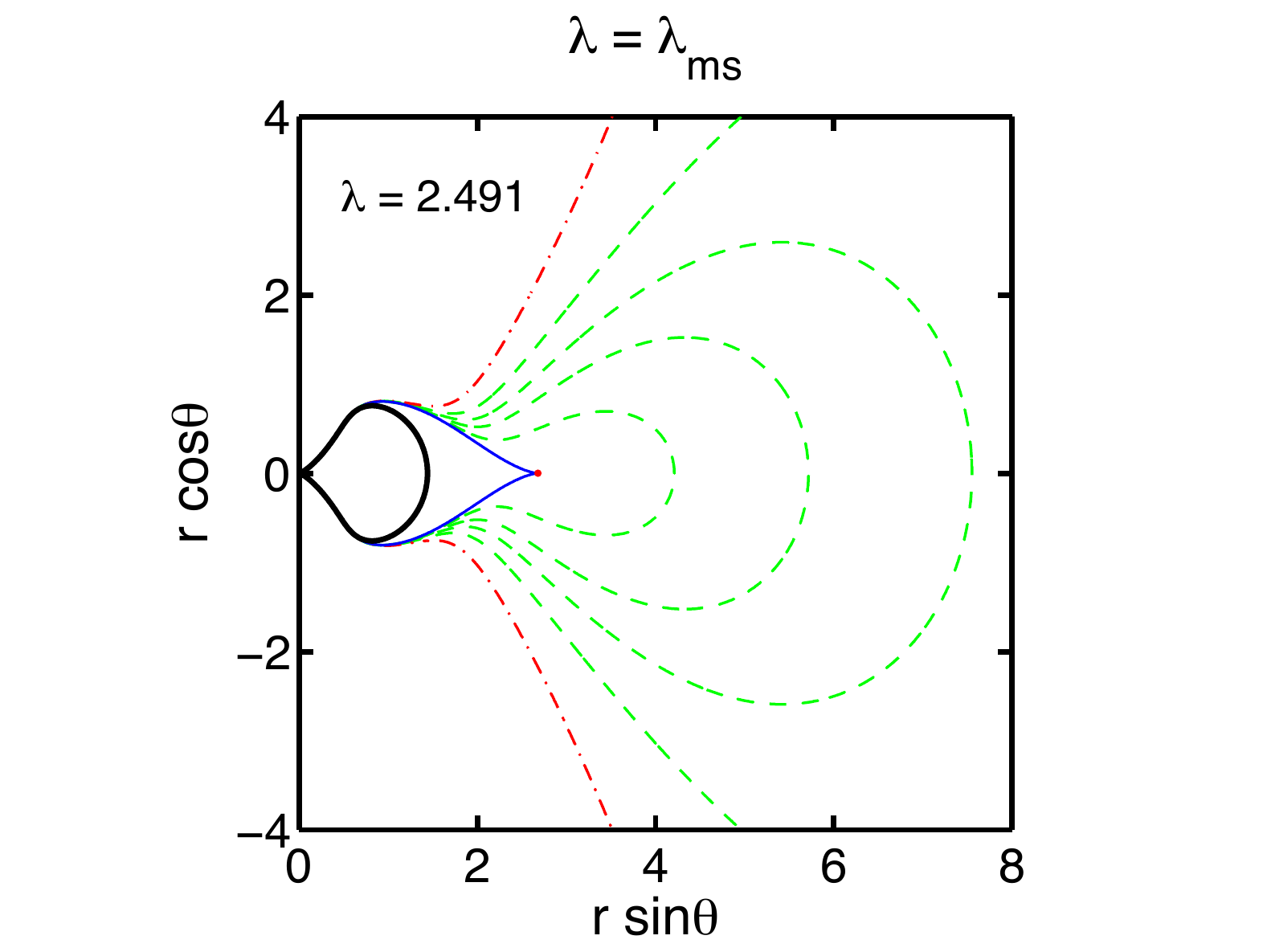} \\ \vspace{0.6cm}
\includegraphics[type=pdf,ext=.pdf,read=.pdf,width=7.6cm]{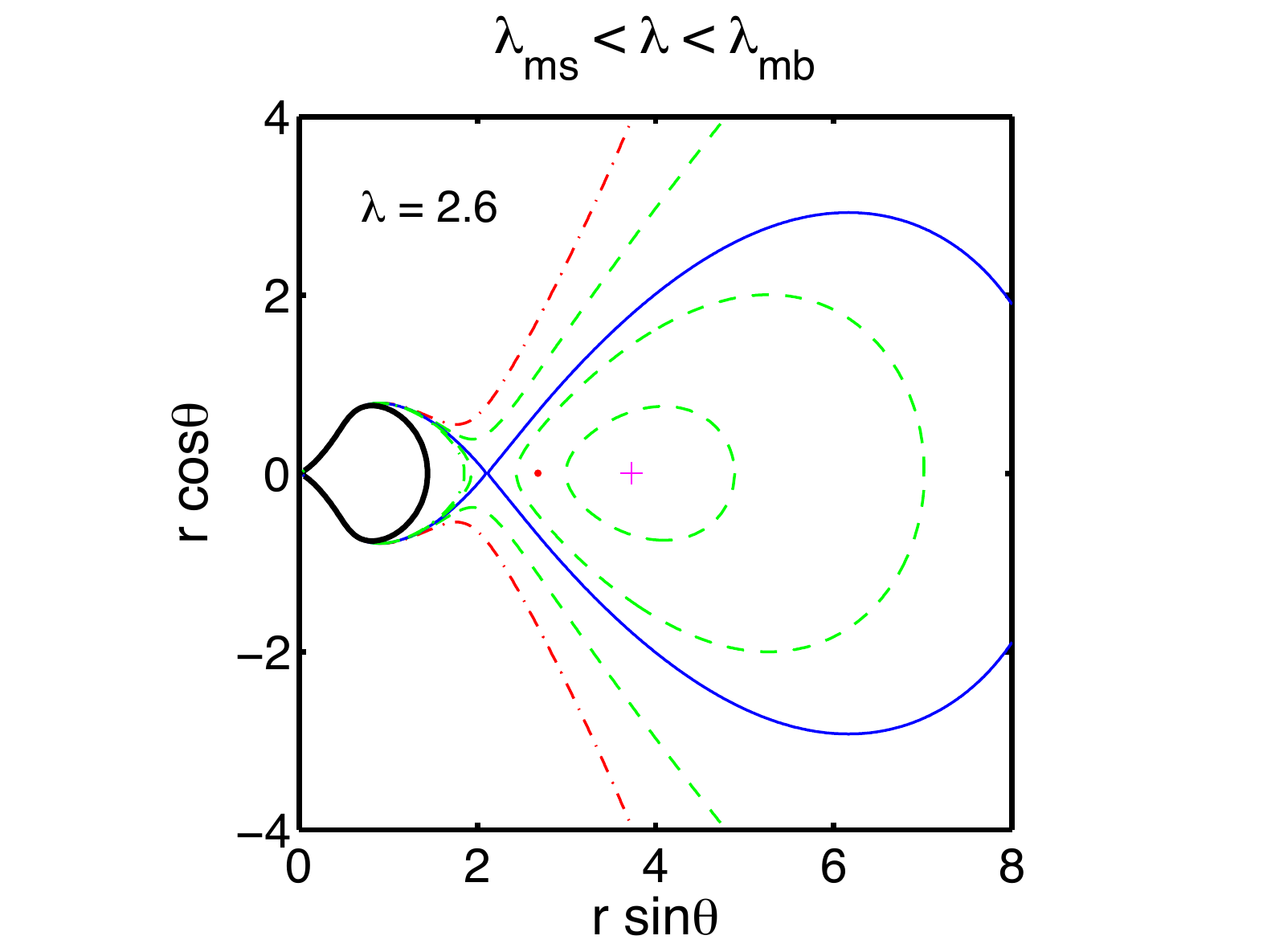}
\includegraphics[type=pdf,ext=.pdf,read=.pdf,width=7.6cm]{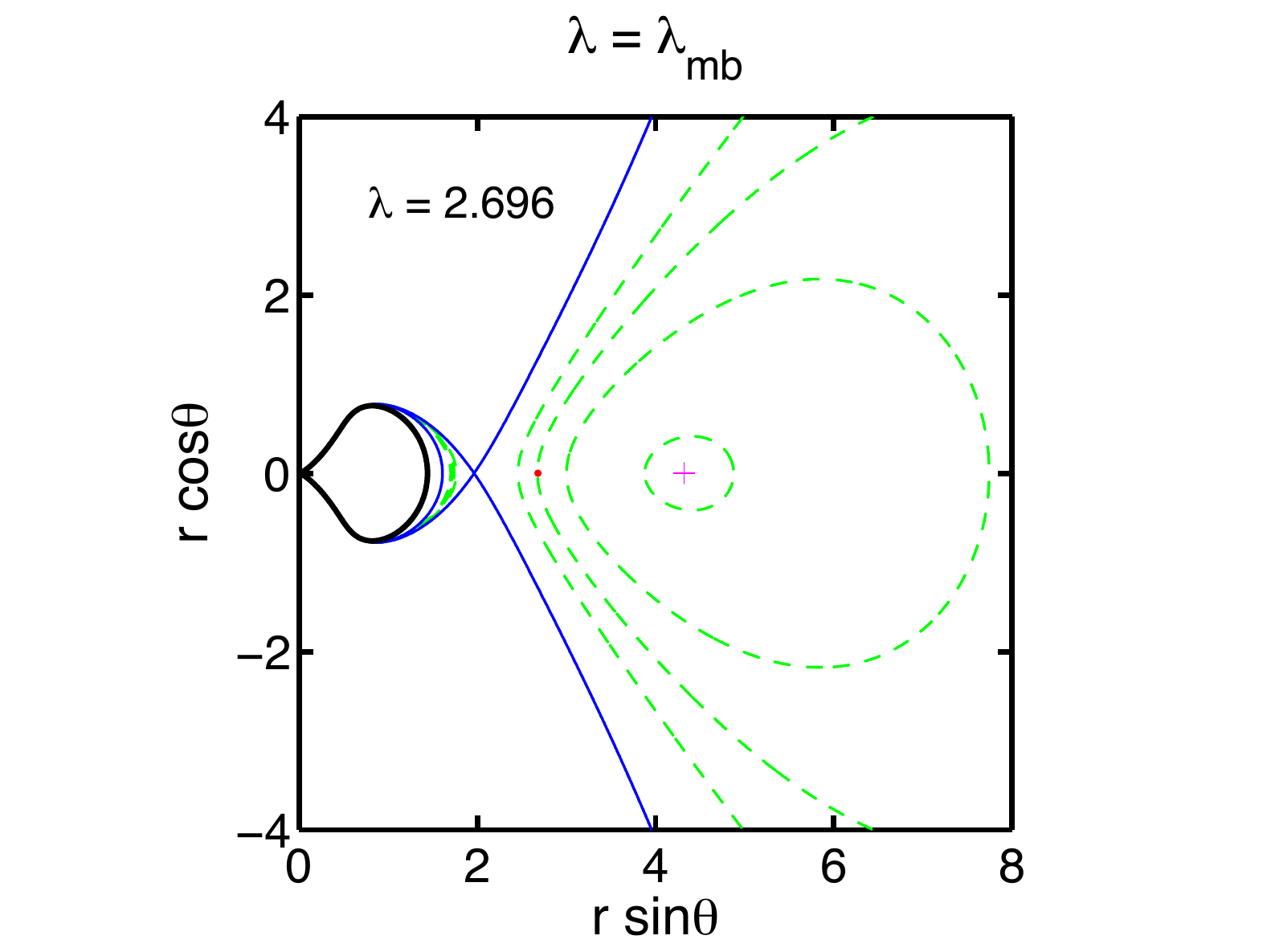} \\ \vspace{0.6cm}
\includegraphics[type=pdf,ext=.pdf,read=.pdf,width=7.6cm]{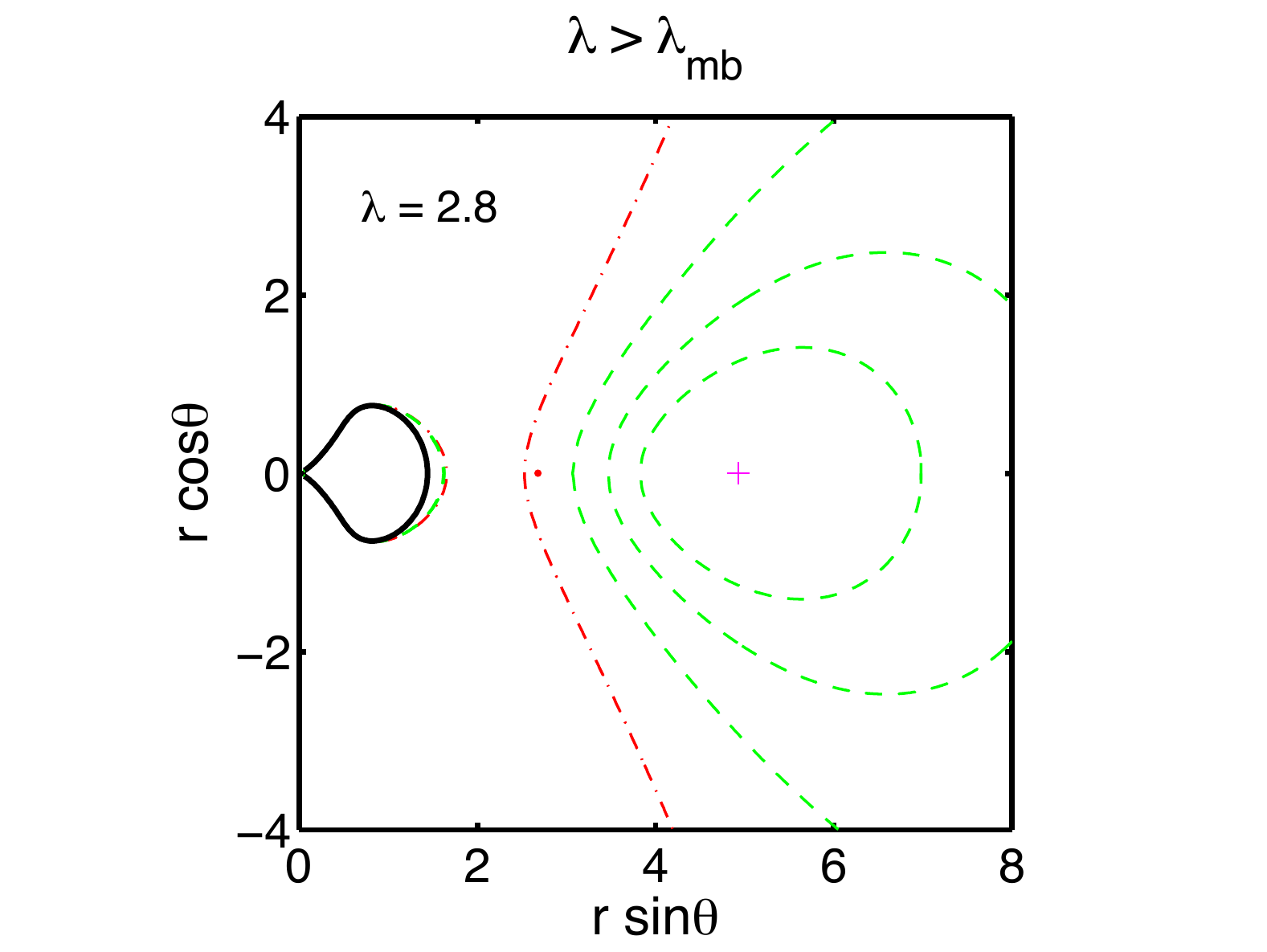}
 \end{center}
\caption{Topology of the equipotential surfaces in the JP background with
$a_* = 1.1$ and $\epsilon_3 = -1$. The black solid line is the BH event horizon;
the solid blue line is the equipotential surface with the cusp (if any); the dashed
green lines are some equipotential surfaces with $W < 0$; the dashed-dotted
red line is the equipotential surface $W = 0$; the magenta cross is the local
minimum of $W$ (the center of the disk with maximal pressure, if any); the red
dot is the location of the marginally stable orbit. Here, $\lambda_{\rm ms} = 2.491$
and $\lambda_{\rm mb} = 2.696$. The picture of the accretion process is
{\it qualitatively similar} to the Kerr case.}
\label{f-1}
\end{figure}

\begin{figure}
\begin{center}
\includegraphics[type=pdf,ext=.pdf,read=.pdf,width=7.6cm]{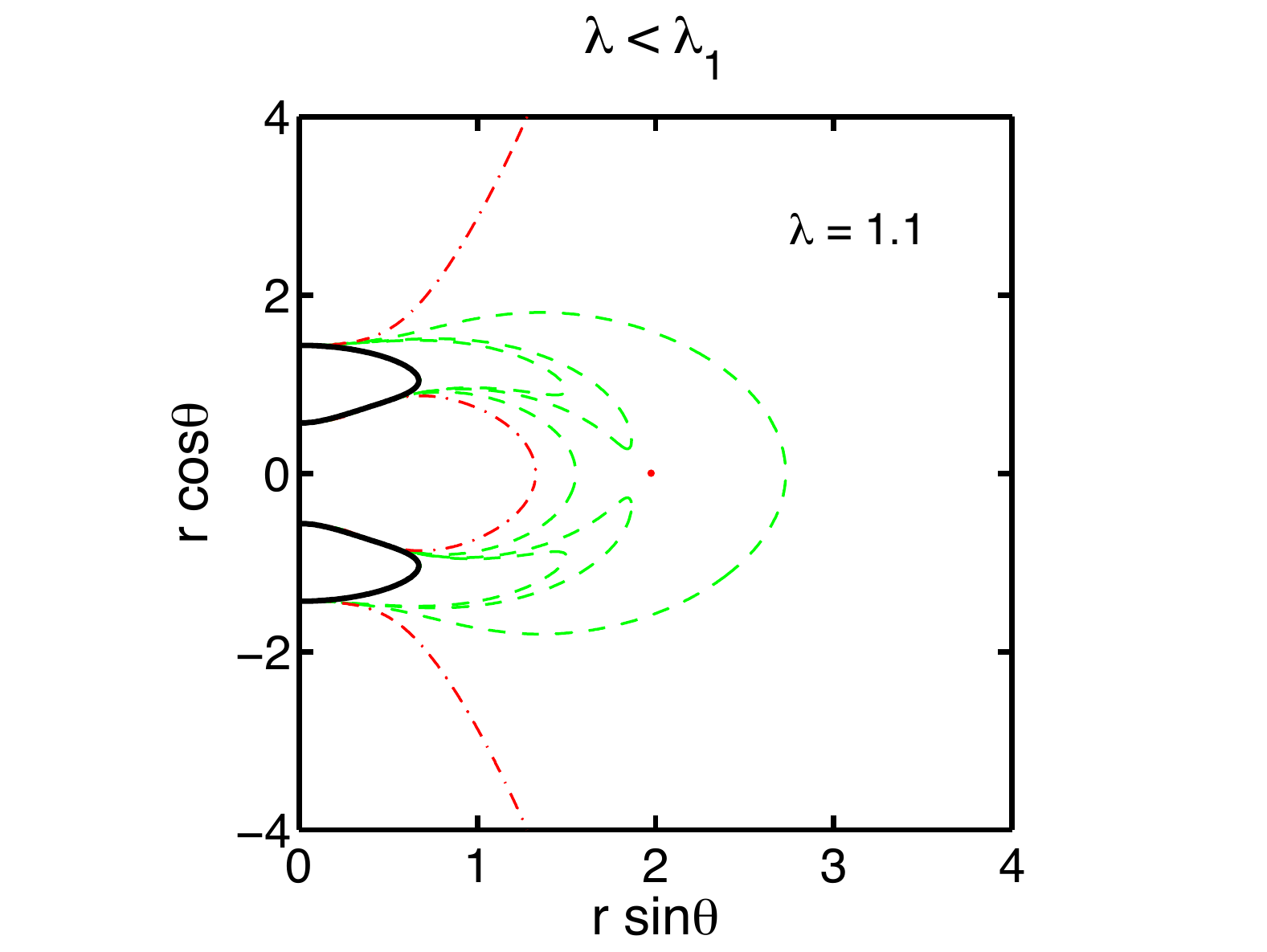}
\includegraphics[type=pdf,ext=.pdf,read=.pdf,width=7.6cm]{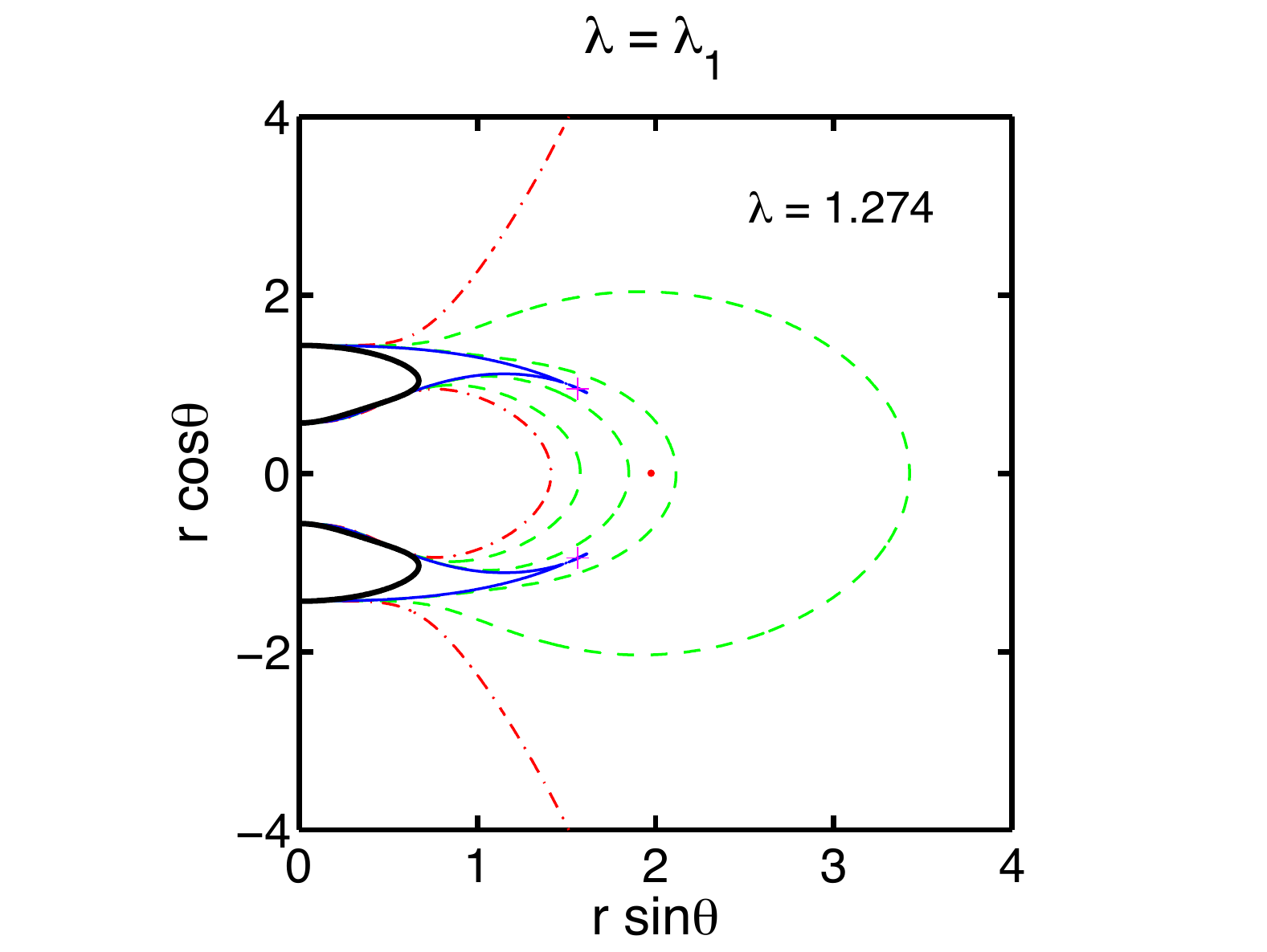} \\ \vspace{0.6cm}
\includegraphics[type=pdf,ext=.pdf,read=.pdf,width=7.6cm]{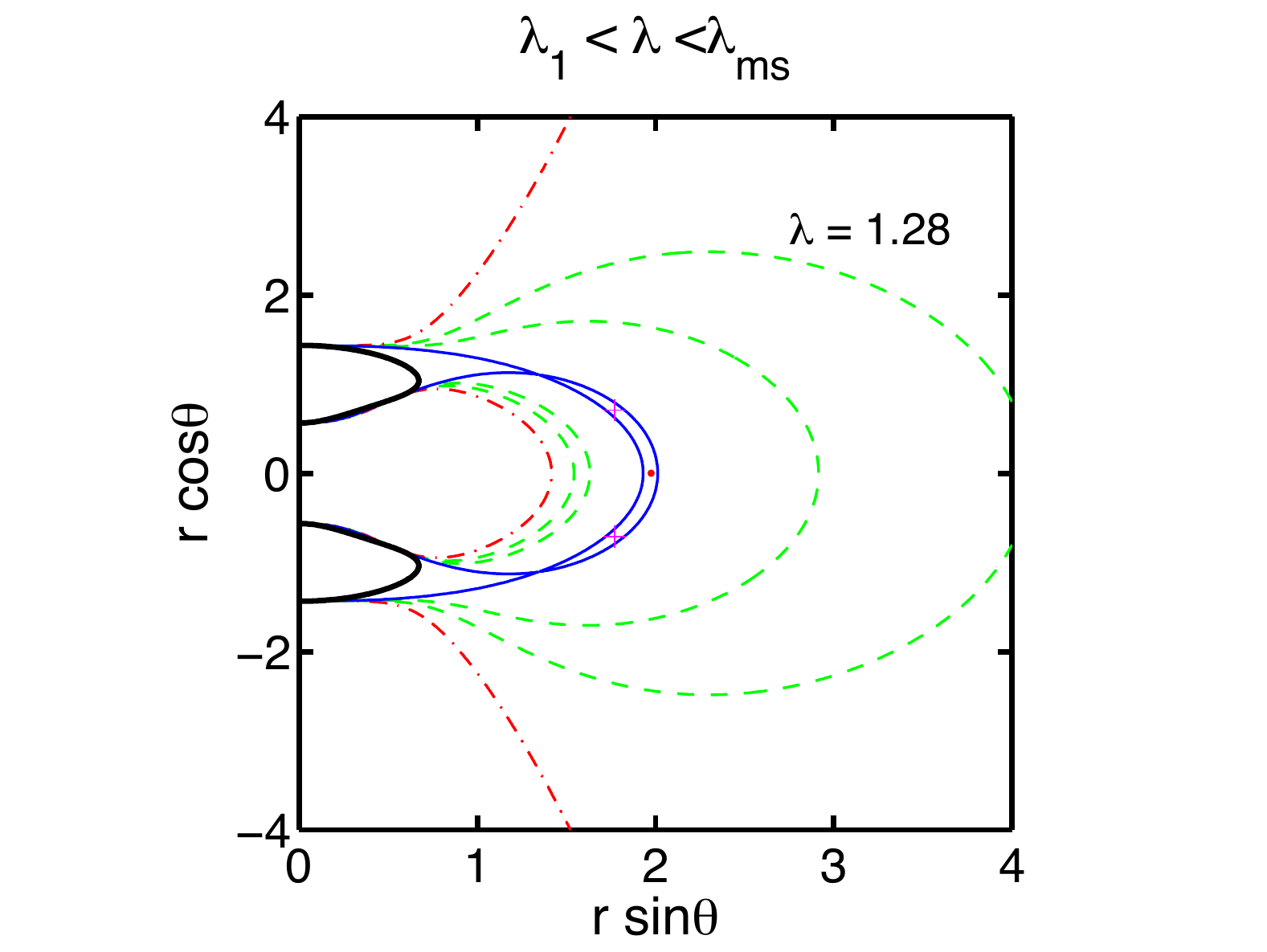}
\includegraphics[type=pdf,ext=.pdf,read=.pdf,width=7.6cm]{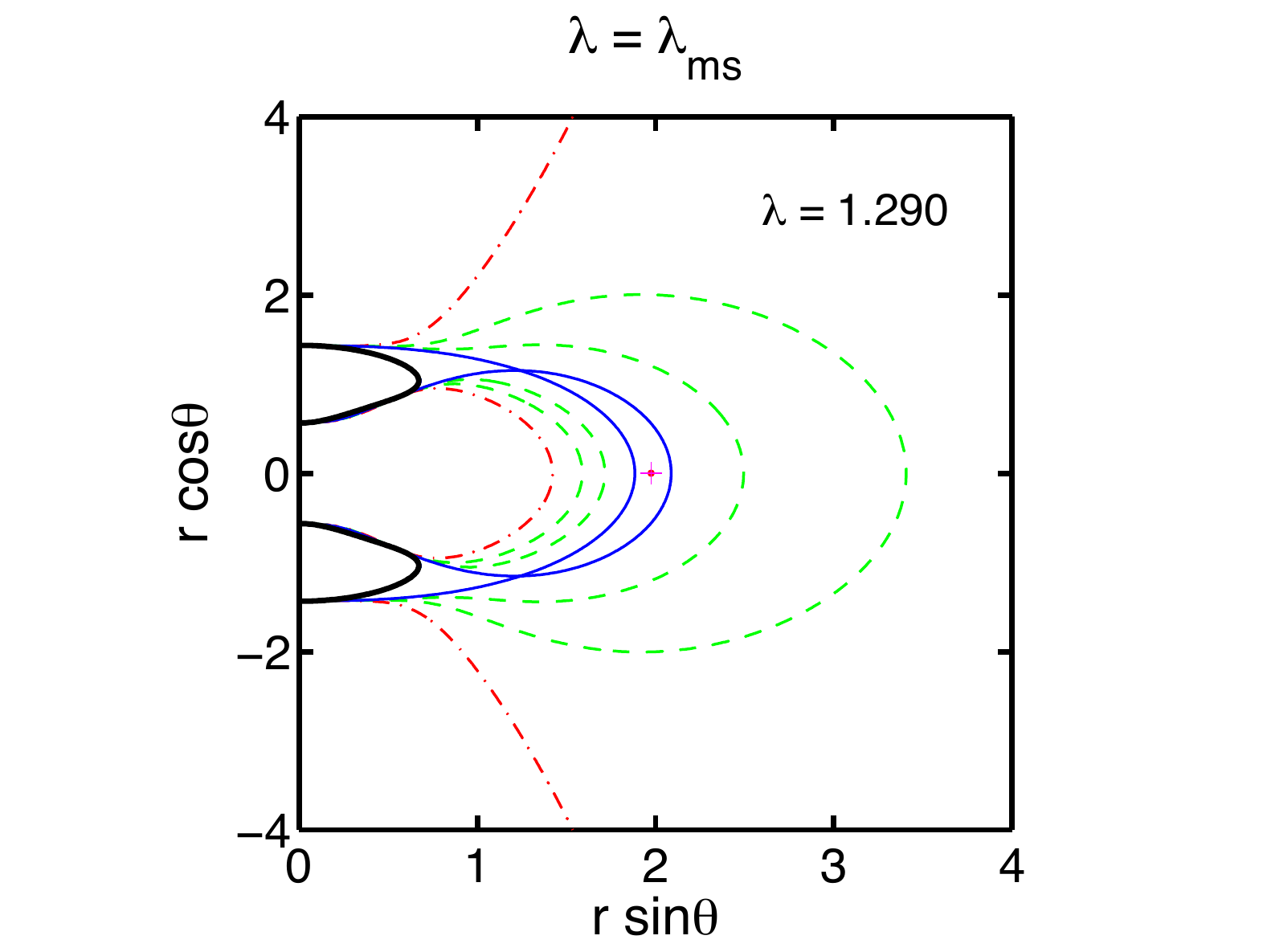} \\ \vspace{0.6cm}
\includegraphics[type=pdf,ext=.pdf,read=.pdf,width=7.6cm]{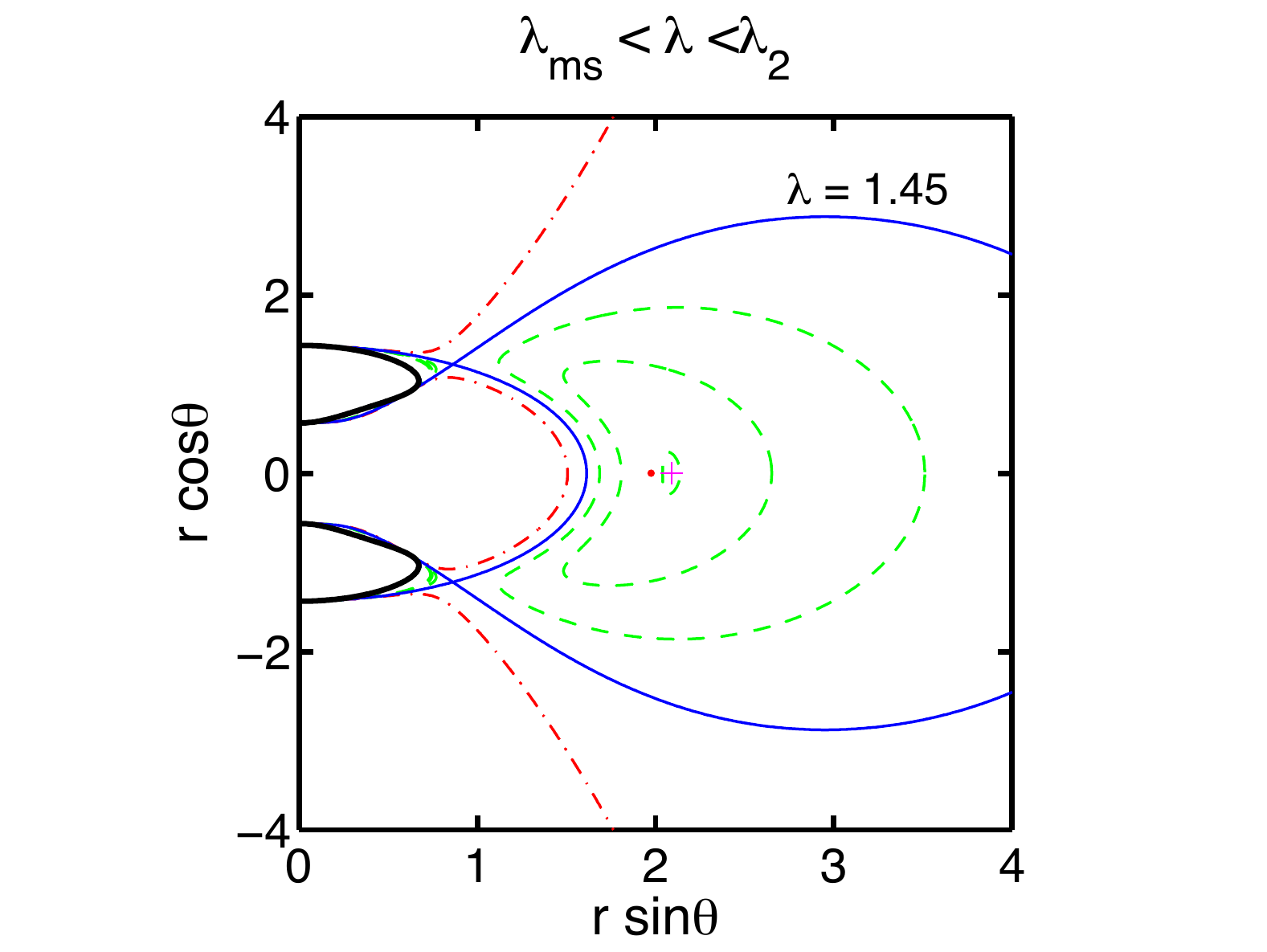}
\includegraphics[type=pdf,ext=.pdf,read=.pdf,width=7.6cm]{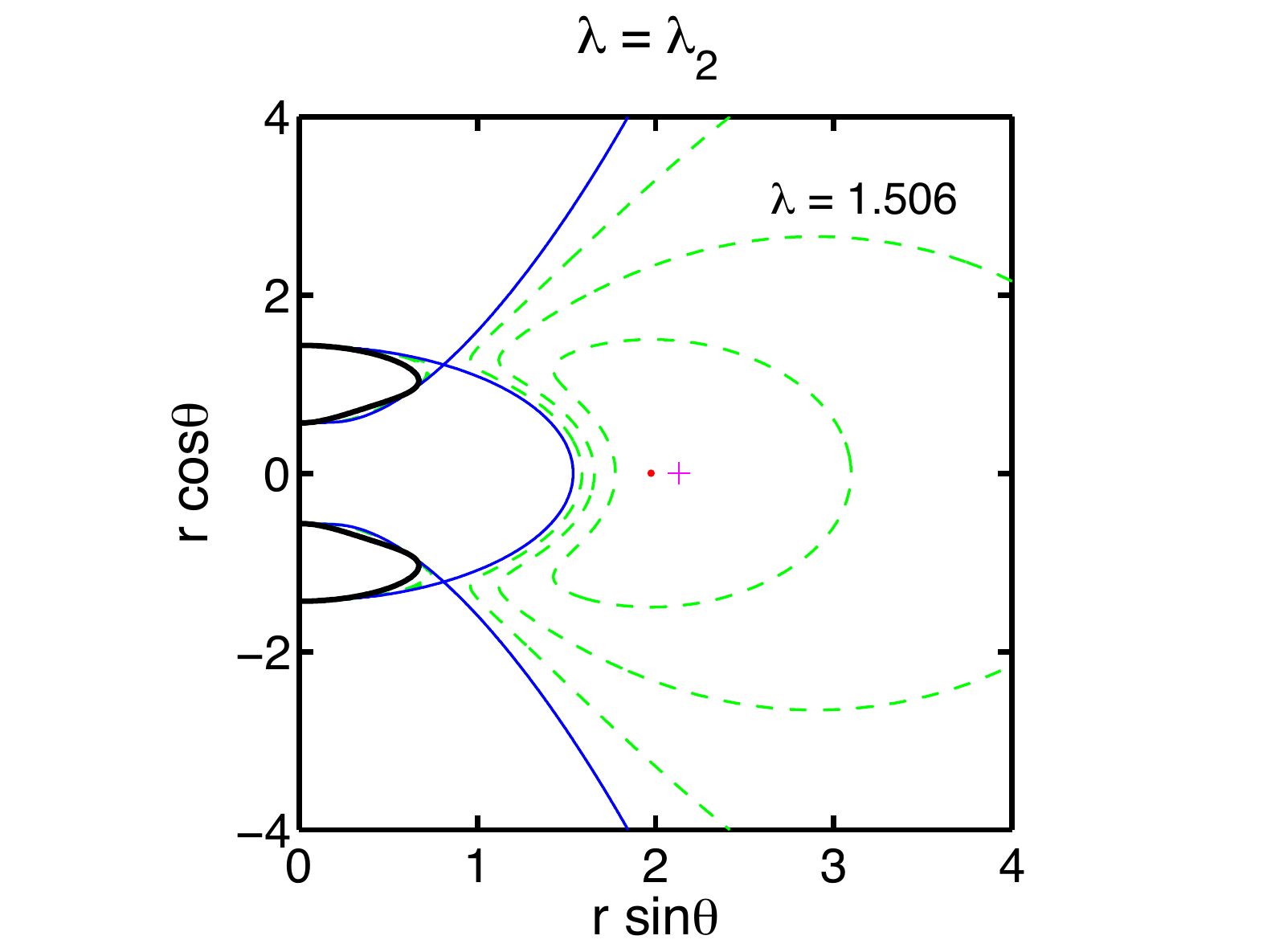}
 \end{center}
\caption{Topology of the equipotential surfaces in the JP background with
$a_* = 0.9$ and $\epsilon_3 = 2$. The black solid line is the BH event horizon;
the solid blue line is the equipotential surface with the cusps (if any); the dashed
green lines are some equipotential surfaces with $W < 0$; the dashed-dotted
red line is the equipotential surface $W = 0$; the magenta crosses are the local
minima of $W$ (the centers of the disk with maximal pressure, if any); the red dot
is the location of the marginally stable orbit. Here, $\lambda_1 = 1.274$,
$\lambda_{\rm ms} = 1.290$, and $\lambda_2 = 1.506$. The picture of the
accretion process is {\it qualitatively different} from the Kerr case.}
\label{f-2}
\end{figure}

\begin{figure}
\begin{center}
\includegraphics[type=pdf,ext=.pdf,read=.pdf,width=8cm]{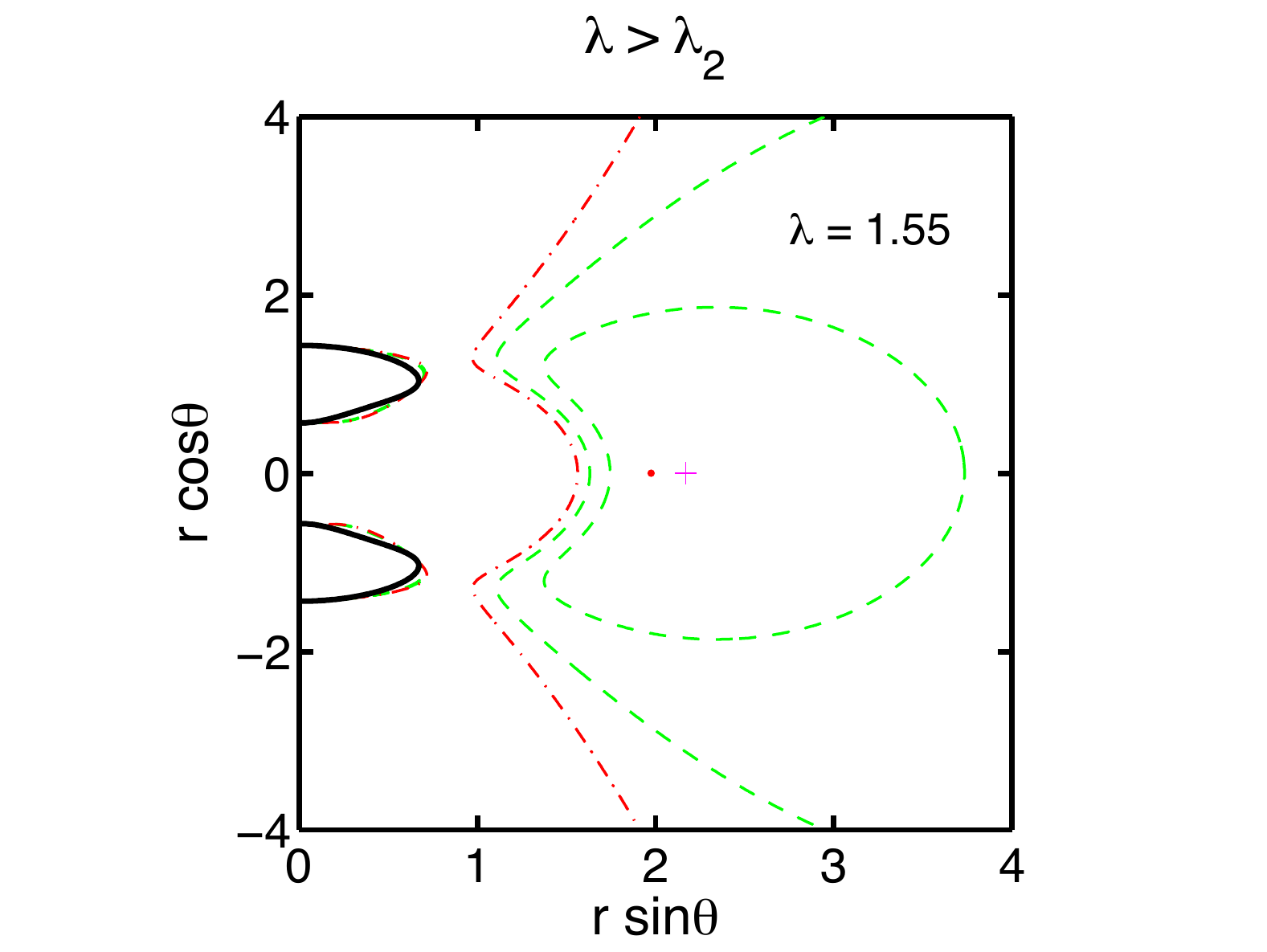}
 \end{center}
\caption{As in Fig.~\ref{f-2} for $\lambda > \lambda_2$.}
\label{f-2b}
\end{figure}

\section{Thick disks in non-Kerr space-times \label{s-nk}}

The Polish doughnut model is formulated for a generic stationary and axisymmetric
space-time. In what follows, we apply this model to two non-Kerr backgrounds.

\subsection{Non-Kerr black holes in an alternative gravity theory\label{ss-jp}}

The Johannsen-Psaltis (JP) metric describes non-Kerr BHs in a putative 
alternative theory of gravity. In its simplest version, the line element in 
Boyer-Lindquist coordinates reads
\be\label{gmn}
ds^2 &=& - \left(1 - \frac{2 M r}{\Sigma}\right) \left(1 + h\right) dt^2
- \frac{4 a M r \sin^2\theta}{\Sigma} \left(1 + h\right) dtd\phi
+ \frac{\Sigma \left(1 + h\right)}{\Delta + a^2 h \sin^2 \theta} dr^2
+ \nonumber\\ &&
+ \Sigma d\theta^2 + \left[ \left(r^2 + a^2 +
\frac{2 a^2 M r \sin^2\theta}{\Sigma}\right) \sin^2\theta +
\frac{a^2 (\Sigma + 2 M r) \sin^4\theta}{\Sigma} h \right] d\phi^2 \, ,
\ee
where $a = a_* M$ and
\be
\Sigma = r^2 + a^2 \cos^2\theta \, , \quad
\Delta = r^2 - 2 M r + a^2 \, , \quad
h = \frac{\epsilon_3 M^3 r}{\Sigma^2} \, .
\ee
The compact object is more prolate (oblate) than a Kerr BH for $\epsilon_3 > 0$
($\epsilon_3 < 0$); when $\epsilon_3 = 0$, we recover the Kerr solution.

In the Kerr background, equatorial circular orbits are always vertically stable,
while they are radially stable only for radii larger than the one of the marginally
stable orbit. In a generic non-Kerr space-times, equatorial circular orbits may 
also be vertically unstable, which leads to a number of new phenomena~\cite{bb,qpo}.
In the case of thick accretion disks in the JP background, we find two qualitatively 
different pictures which are related to the radial or vertical instability of the 
marginally stable orbit, even if the case of the transition space-times is less clear.

The ISCO is radially marginally stable when the central objects is a Kerr BH
($\epsilon_3 = 0$), when it is more oblate than a Kerr BH ($\epsilon_3 < 0$), and
also when it is more prolate than a Kerr BH and the spin parameter is below
a critical value that depends on the deformation parameter ($\epsilon_3 > 0$
and $a_* < a_*^{\rm crit} (\epsilon_3)$). The accretion process proceeds as
described in the previous section and it is set by the value of $\lambda$ with
respect to the ones of $\lambda_{\rm ms}$ and $\lambda_{\rm mb}$. The
topology of the equipotential surfaces in the JP background with $a_* = 1.1$
and $\epsilon_3 = -1$ is shown in Fig.~\ref{f-1}. There are no qualitative differences
with the Kerr case (see Fig.~3 in~\cite{polish2}).

The second case, in which the marginally stable orbit is vertically unstable, occurs
when the compact object is more prolate than a Kerr BH and its spin parameter
is above a critical value ($\epsilon_3 > 0$ and $a_* > a_*^{\rm crit} (\epsilon_3)$).
Now, the space-time has no marginally bound orbit, as the marginally stable orbit
is not a minimum of the particle energy. So, $\lambda_{\rm mb}$ cannot be defined.
In addition to $\lambda_{\rm ms}$, we find other two relevant values, say
$\lambda_1$ and $\lambda_2$, with $\lambda_1 < \lambda_{\rm ms} < \lambda_2$.
There are thus seven qualitatively different scenarios:
\begin{enumerate}
\item $\lambda < \lambda_1$ (Fig.~\ref{f-2}, top left panel). No disks are possible,
as there are no closed equipotential surfaces.
\item $\lambda = \lambda_1$ (Fig.~\ref{f-2}, top right panel). There are two
local minima of $W$, above and below the equatorial plane, corresponding to the
two disks' centers. They are not really two minima, but two flexes. The disks exist
as two infinitesimally thin unstable rings outside the equatorial plane.
\item $\lambda_1 < \lambda < \lambda_{\rm ms}$ (Fig.~\ref{f-2}, central left panel).
There are two local minima of $W$, above and below the equatorial plane.
We may have either two non-equatorial toroidal disks or one disk crossing the
equatorial plane. There is one configuration with two cusps, which are located
above and below the equatorial plane.
\item $\lambda = \lambda_{\rm ms}$ (Fig.~\ref{f-2}, central right panel).
$W$ has a local minimum located at the marginally stable orbit on the equatorial
plane. There are many stationary configurations without cusps and one with
two cusps, which are located above and below the equatorial plane.
\item $\lambda_{\rm ms} < \lambda < \lambda_2$ (Fig.~\ref{f-2}, bottom left panel).
$W$ has a local minimum located on the equatorial plane at a radius larger than
the one of the marginally stable orbit. There are many stationary configurations
without cusps and one with two cusps, which are located above and below the
equatorial plane.
\item $\lambda = \lambda_2$ (Fig.~\ref{f-2}, bottom right panel). The equipotential
surface with the two cusps is the one with $W = 0$. Accretion is possible in the
limit of a disk of infinite size.
\item $\lambda > \lambda_2$ (Fig.~\ref{f-2b}). No accretion is possible, as there
are no equipotential surfaces $W \le 0$ with cusps.
\end{enumerate}
Unfortunately, $\lambda_1$ and $\lambda_2$ are found numerically and, unlike
$\lambda_{\rm ms}$ and $\lambda_{\rm mb}$, there is apparently not a simple 
interpretation of them in term of the $\lambda$ of a free particle. 
Let us also note that, as pointed out in Refs.~\cite{topo0,topo1,topo2}, the topology
of the event horizon of these BHs may be non-trivial. In Fig.~\ref{f-1}, the BH
horizon has the topology of a torus. In Figs.~\ref{f-2} and \ref{f-2b}, it has the
topology of two 2-spheres.

\subsection{Exotic compact objects without horizon in General Relativity \label{ss-mms}}

The Manko-Mielke-Sanabria-G\'omez (MMS) metric~\cite{mms} is an exact solution 
of the Einstein-Maxwell equations. It can describe the exterior gravitational field 
around an exotic compact object. The equation of state of ordinary matter cannot 
explain very compact objects exceeding 3~$M_\odot$ and for this reason here 
we include the word ``exotic''. We consider the version with three parameters: 
the mass $M$, the specific spin angular momentum $a = J/M$, and the parameter 
$b$. In Ref.~\cite{mms}, the metric was written in prolate spheroidal coordinates, 
which are suitable only for slow-rotating compact objects. It was extended to 
objects with spin parameter larger than 1 in Ref.~\cite{spin1b}, by a coordinate 
transformation to oblate spheroidal coordinates. As in the next section we are 
interested in fast-rotating objects, here we write the metric in oblate spheroidal 
coordinates and we refer to the original paper for the slow-rotating case. The 
line element is
\be
ds^2 &=& - f \left(dt - \omega d\phi\right)^2
+ \frac{k^2 e^{2\gamma}}{f}\left(x^2 + y^2\right)
\left(\frac{dx^2}{x^2 + 1} + \frac{dy^2}{1 - y^2}\right) 
+ \nonumber\\ &&
+ \frac{k^2}{f} \left(x^2 + 1\right)\left(1 - y^2\right) d\phi^2 \, ,
\ee
where 
\be
f = \frac{A}{B} \, , \quad
\omega = - (1 - y^2) \frac{C}{A} \, , \quad
e^{2\gamma} = \frac{A}{16 k^8 (x^2 + y^2)^4} \, .
\ee
$k = \sqrt{-d-\delta}$ and
\be
\delta = -\frac{M^2 b^2}{M^2 - (a-b)^2} \, , \qquad
d = \frac{M^2 - (a-b)^2}{4} \, .
\ee
The functions $A$, $B$, and $C$ can be written in the following 
compact way~\cite{mms}
\be
A = R^2 + \lambda_1 \lambda_2 S^2 \, , \quad
B = A + R P + \lambda_2 S T \, , \quad
C = R T - \lambda_1 S P \, .
\ee
Here $\lambda_1 = k^2 (x^2 + 1)$, $\lambda_2 = y^2 - 1$, and
\be
P &=& 2kMx[(2kx + M)^2 - 2y^2(2\delta + ab - b^2) - a^2 + b^2]
- 4y^2(4\delta d - M^2 b^2) \, , \nonumber\\
R &=& 4[k^2(x^2+1)+\delta(1-y^2)]^2+(a-b)
[(a-b)(d-\delta) - M^2b](1-y^2)^2\, , \nonumber\\
S &=& -4\{(a-b)[k^2(x^2+y^2)+2\delta y^2]+y^2M^2b\} 
\, , \nonumber\\
T &=& 8Mb(kx + M)[k^2(x^2 + 1) + \delta(1 - y^2)] +
\nonumber\\ &&
+ (1 - y^2)\{(a - b)(M^2b^2 - 4\delta d)
- 2M(2kx + M)[(a-b)(d-\delta) - M^2 b]\} \, .
\ee 
The Kerr metric is recovered for $b = \sqrt{a^2 - M^2}$, so $b$ is not the usual 
deformation parameter which measures deviations from the Kerr geometry.

As in the case of the JP space-time, the hydrodynamical structure of thick 
accretion disks in the MMS background can be immediately obtained by using
the corresponding metric coefficient. The qualitative picture is the same we
have already discussed in the JP space-time: we have the Kerr-like scenario
of Fig.~\ref{f-1} when the ISCO is set by the orbital stability along the radial 
direction, and the two-cusps scenario of Figs.~\ref{f-2} and \ref{f-2b} when the 
ISCO is marginally vertically stable. At this point, a quantitative comparison of 
the two metrics is not possible, as the parameters $\epsilon_3$ and $b$ have
not a clear physical meaning. However, as discussed in the next sections, 
when we consider observable phenomena, the physically relevant properties
are quite similar, despite the completely different origin of the two space-times.

\begin{figure}
\begin{center}
\includegraphics[type=pdf,ext=.pdf,read=.pdf,width=7cm]{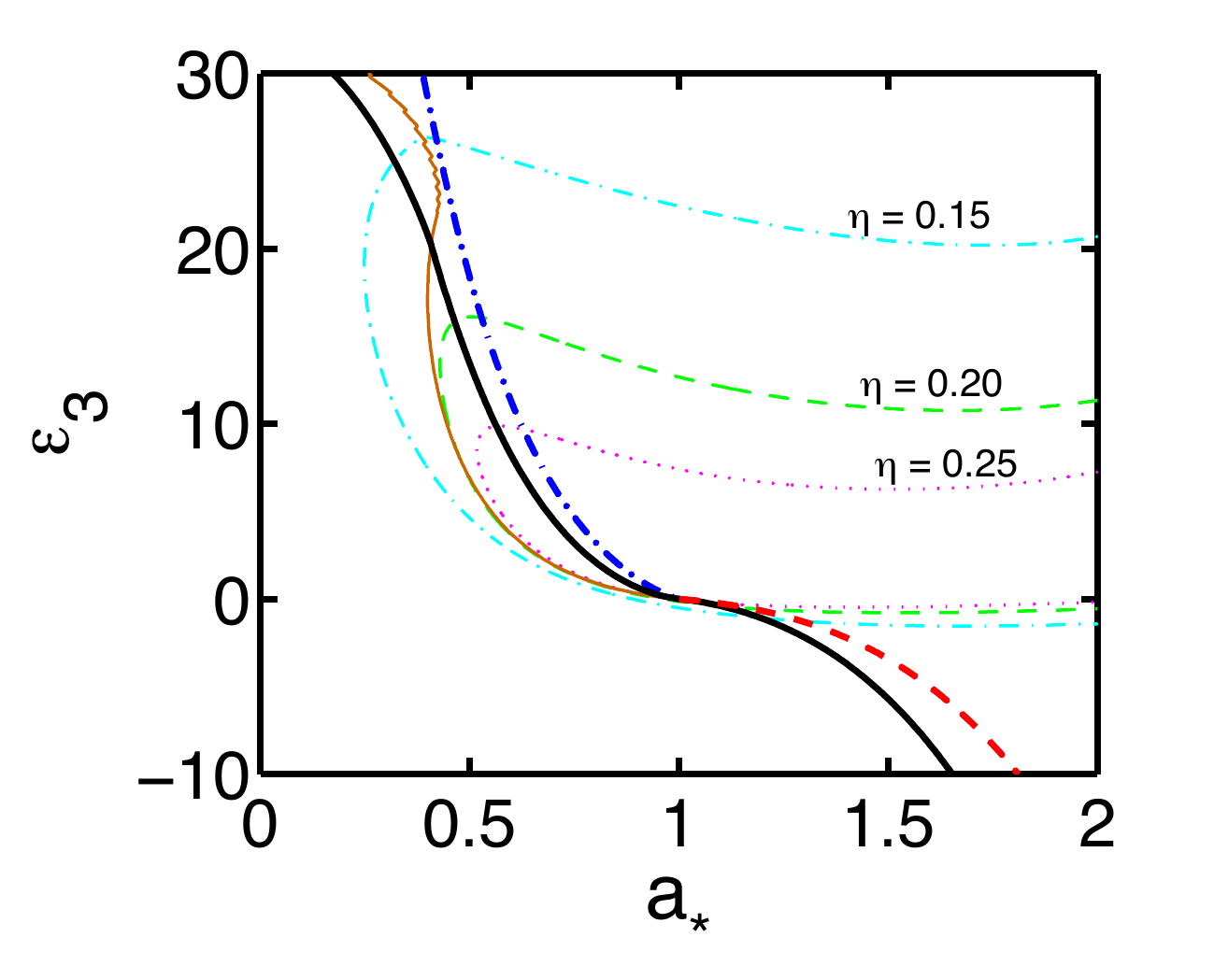}
\includegraphics[type=pdf,ext=.pdf,read=.pdf,width=6.9cm]{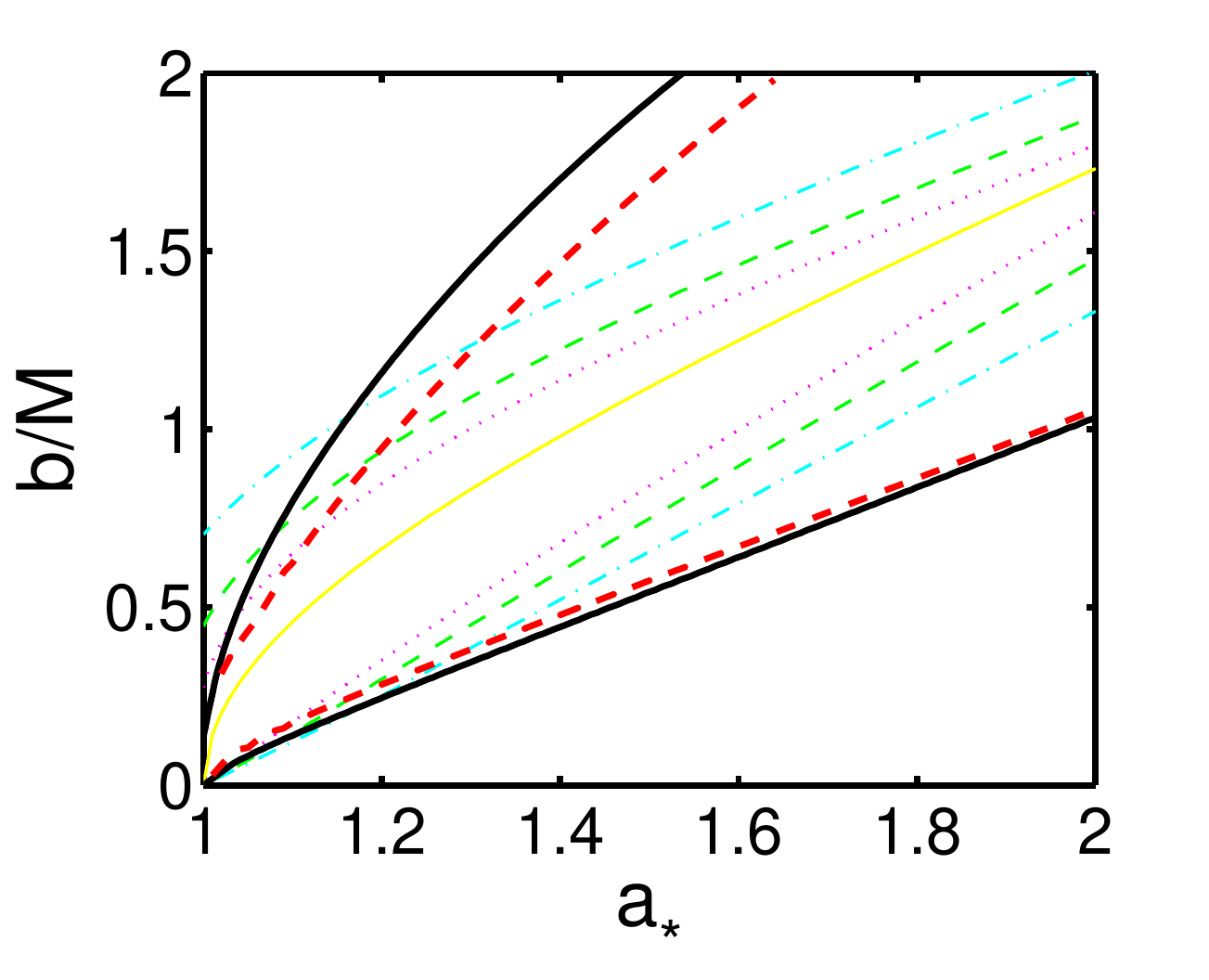}
 \end{center}
\caption{Spin parameter-deformation parameter plane. Curve of $a_*^{\rm eq}$
from accretion from a thick disk (thick dashed red curve for $\lambda = \lambda_{\rm mb}$
and thick dashed-dotted blue curve for $\lambda = \lambda_2$), curve of
$a_*^{\rm eq}$ from accretion from a thin disk (thick solid black curve),
Novikov-Thorne radiative efficiency $\eta_{\rm NT} = 0.15$ (thin dashed-dotted light
blue curve), 0.20 (thin dashed green curve), and 0.25 (thin dotted magenta curve).
The thin solid brown curve (left panel) for $\epsilon_3 > 0$ separates the regions of the plane
in which the ISCO is marginally vertically (right side) and marginally radially (left side)
stable. For $b = M \sqrt{a_*^2-1}$ (thin solid yellow curve in the right panel), we recover 
the Kerr solution. Left panel: JP background. Right panel: MMS background.}
\label{f-3}
\end{figure}

\section{Evolution of the spin parameter \label{s-spin}}

An accreting compact object changes its mass and spin angular momentum as
a consequence of the accretion process. Solid results of the spin evolution would
necessarily require realistic non-stationary models of disks accretion and jet
emission. That is not possible at present, and therefore we can just estimate
the spin evolution with simple (or very simple) accretion models, whose results
have to be taken with some caution. The general picture is that the accreting gas
falls to the compact object by loosing energy and angular momentum. When it
reaches the inner edge of the disk, it plunges onto the compact object. If the
gas emits no or negligible additional radiation after having plunged, the
mass and the spin of the compact object change by $\delta M = - u_t^{\rm in}
\delta m$ and $\delta J = u_\phi^{\rm in} \delta m$, where $- u_t^{\rm in}$ and
$u_\phi^{\rm in}$ are, respectively, the specific energy and the specific angular
momentum of the gas's particles at the inner edge of the disk, while $\delta m$ is
the gas rest-mass. The evolution of the spin parameter is thus governed by the
following equation:
\be
\frac{da_*}{d\ln M} = \frac{\lambda_{\rm in}}{M} - 2 a_* \, .
\ee

In the case of a thin accretion disk in the Kerr background, $\lambda_{\rm in}$
is supposed to be $\lambda_{\rm ms}$ (Novikov-Thorne model) and the
equilibrium value of the spin parameter is $a_*^{\rm eq} = 1$~\cite{bard}; that is,
the object is spun up by the accretion process if $a_* < 1$, and spun down if
$a_* > 1$. Including the effect of the radiation emitted by the disk and captured
by the BH, one finds the famous Thorne's limit $a_*^{\rm eq} = 0.9978$ (when
the disk's emission is assumed isotropic)~\cite{th}, as the radiation with angular
momentum opposite to the BH spin has larger capture cross-section. The effect
of the returning radiation (the radiation emitted by the disk returning to the disk
as a consequence of light bending) introduces a smaller correction to the Thorne's
limit and the equilibrium spin parameter changes to $a_*^{\rm eq} = 0.9983$~\cite{li05}.

In the case of thick disks, one can note that in the Polish doughnut model the inner
edge of the disk is inside the marginally stable orbit and it can be arbitrary close
to the marginally bound one: here the radiative efficiency $\eta = 1 + u_t^{\rm in}$
goes to zero and therefore the spin evolution is very weakly affected by the
emission of radiation, suggesting that the Thorne's bound can be exceeded~\cite{lasota}.
However, a more detailed calculation requires to include the effect of the fluid's
viscosity, which was assumed to be completely negligible in~\cite{lasota}. Very recently,
in Ref.~\cite{sado} the authors have studied the spin evolution for high accretion
rates in a relativistic, advective, optically thick slim accretion disk model. They
find that the BH spin evolution is hardly affected by the emitted radiation at high
accretion rate ($\dot{M} \gtrsim 10 \dot{M}_{\rm Edd}$, where $\dot{M}_{\rm Edd}$ is
the Eddington mass accretion rate) and that the equilibrium spin value is determined
by the flow properties. For $\dot{M} = 10 \dot{M}_{\rm Edd}$ and a viscosity parameter
$\alpha = 0.01$, they get $a_*^{\rm eq} = 0.9994$~\cite{sado}.

The spin evolution in non-Kerr space-times (non-Kerr BHs and compact objects
without an event horizon) have been discussed in~\cite{spin1a,spin1b,spin2,spin3},
but only in the case of accretion from thin disk. The key-result of those papers is 
that the bound $|a_*| \le 1$ can be violated. 
With the results of the previous section, we can here consider the case of thick
accretion disks. The Polish doughnut model, in which the viscosity parameter $\alpha$
is supposed to be completely negligible, is surely an extremely simplified model.
Moreover, we have studied the case of marginally stable disks with $\lambda = {\rm const.}$,
which is known to be unstable on a dynamical time-scale (a few orbital
periods)~\cite{runaway1}. Nevertheless, we believe that our simple prescription
can provide the correct insight. These accretion disks become stable for very small
values of the angular momentum slope~\cite{runaway2}. As shown in~\cite{lasota} in
the Kerr background, even here the effect of the emitted radiation should become
negligible for super-Eddington rates\footnote{Let us note that, even when
$\lambda \rar \lambda_2$, the radiative efficiency $\eta = 1 + u^{\rm in}_t \rar 0$
and therefore the emitted radiation is likely irrelevant.}.
In the end, our goal is to find a conservative upper limit
to the maximum value of the spin parameter of these objects and all our
approximations can only overestimate $a_*^{\rm eq}$, so our final result is
indeed conservative. The equilibrium values of the spin parameter
as a function of the deformation one are reported in Fig.~\ref{f-3}, respectively
for the JP (left panel) and MMS (right panel) backgrounds.

\section{Discussion \label{s-dd}}

As discussed in~\cite{spin2}, the value of the spin parameter of a compact object
is determined by the competition of three physical processes: the event creating
the object, mergers, and gas accretion. In the case of the super-massive BH
candidates at the centers of galaxies, the value of the natal spin is completely irrelevant,
as the object has increased its mass by several orders of magnitude and the
spin has changed accordingly. Minor mergers and short-term accretion events
are random events and the net effect is to spin the compact object down.
Major mergers can unlikely produce very-fast rotating objects and therefore
the maximum value of the today spin parameter of the super-massive BH
candidates in galactic nuclei is likely $a_*^{\rm eq}$, which is reached after long-term
accretion from a disk. Let us note that this is true even if the initial spin parameter is
higher than $a_*^{\rm eq}$, as in this case the long-term accretion processes
(and, even more efficiently, the other accretion mechanisms) would spin the
compact object down to $a_*^{\rm eq}$. With this observation, we can exclude 
the region on the right hand side of the curve $a_*^{\rm eq}$ on the plane 
spin-deformation parameter, see Fig.~\ref{f-3}.

The left hand side of the spin-deformation parameter plane can be constrained by
current spin measurements, which are usually obtained by assuming the Kerr
background. At present, the two most popular and reliable approaches are the
continuum-fitting method (see e.g.~\cite{mcc,refe-cfm1,refe-cfm2} and references 
therein) and the analysis of the K$\alpha$ iron line (for a review, see e.g.~\cite{refe-ka}).
The continuum-fitting method infers the spin value by studying the thermal 
spectrum of a thin accretion disk. As discussed in Refs.~\cite{b1,b3,b4,test2}, 
actually this technique measures the radiative efficiency of the Novikov-Thorne
model, $\eta_{\rm NT} = 1 + u_t^{\rm ms}$. It is thus easy to translate a spin
measurement under the Kerr assumption into an allowed region on the spin-deformation
parameter plane. For instance, the spin of the stellar-mass BH candidate in
GRS~1915+105 has been estimated $a_* > 0.98$ in~\cite{grs}. Such a measurement
corresponds to $\eta_{\rm NT} > 0.234$ in terms of the radiative efficiency of the 
Novikov-Thorne model. However, the continuum-fitting method can be applied
only to stellar-mass BH candidates: the temperature of the disk is proportional
to $M^{-1/4}$ and therefore the disk's peak is around 1~keV for an object with
$M \sim 10$~$M_\odot$, but in the UV range for a super-massive BH candidate.
In the latter case, dust absorption prevents a spin measurement. On the contrary,
our argument is true only for super-massive BH candidates. In the case of
stellar-mass objects in X-ray binaries, the spin more likely reflects the value at
the time of formation of the BH candidate. The accretion disk originates from the
material stripped from the stellar companion. If the latter is massive 
($M \gtrsim 10$~$M_\odot$), its lifetime is too short to alter significantly the mass
and the spin of the BH candidate, even assuming an accretion rate at the 
Eddington limit. If the stellar companion has a mass $M \sim M_\odot$, even after
swallowing the whole star the mass and the spin of the BH candidate cannot 
change significantly.

The study of the shape of the K$\alpha$ iron line can instead probe the space-time
geometry of both stellar-mass and super-massive BH candidates. In the case
of the super-massive BH candidates, there are at least two objects that seem to 
rotate quite fast: for both MGC-6-30-15 and 1H~0707-495, the spin has been
estimated $a_*>0.98$, respectively in Ref.~\cite{refe-a1} and Ref.~\cite{refe-a2}.
The information on the space-time geometry in the K$\alpha$ iron line have
been discussed quite in detail in Ref.~\cite{b6}. Unlike the continuum-fitting method,
in this case it seems there is no simple recipe to translate a spin measurement
under the Kerr background hypothesis into an allowed region on the
spin-deformation parameter plane. Despite that, we can anyway say that the
correlation between spin and deformation parameter in the profile of the
K$\alpha$ iron line is not too different from the one of the disk's thermal spectrum.
So, the combination of the spin measurements from the continuum-fitting method
and the K$\alpha$ iron line of the same object can test the Kerr-nature of a BH
candidate only if the two measurements are quite good (for this purpose, the use
of the jet power discussed in~\cite{b3,b4} would be more efficient, but the 
problem is that the exact mechanism responsible for the production of jets in not 
yet clear). For instance, in Ref.~\cite{refe-cfmka}, the authors consider the stellar-mass
BH candidate XTE~J1550-564. They find $a_* = 0.34$ ($-0.11<a_*<0.71$ at 
90\% C.L.) from the continuum-fitting method, and $a_* = 0.55^{+0.15}_{-0.22}$
from the analysis of the relativistic K$\alpha$ iron line. In Ref.~\cite{b6}, one of
us considered as example the K$\alpha$ iron line generated around a JP BH with 
$a_* = 0.20$ and $\epsilon_3 = 7$, which is a quite deformed object. It turns out 
that such a JP BH has a thermal spectrum indistinguishable from a Kerr BH with
$a_* = 0.68$ and a K$\alpha$ iron line indistinguishable form a Kerr BH with
$a_* = 0.47$ (see section~IV.B and Fig.~9 of Ref.~\cite{b6}). Here for 
``indistinguishable'' we do not mean it cannot be distinguished by current
X-ray measurements, but that the difference is really negligible and indeed
impossible to detect even in the foreseeable future. So, the combination of two
different techniques is surely the only way to test the Kerr-nature of BH candidates,
and the combination of the continuum-fitting method and of the K$\alpha$ iron
line is likely the best option for the near future, but current measurements are
not really good enough to put interesting constraints.

A third method potentially capable of providing a mean value of the spin 
parameter of super-massive BH candidates is based on the Soltan's 
argument~\cite{soltan}, which relates the mean BH mass density with 
the mean energy density radiated by super-massive BHs in the current 
Universe. The argument provides an estimate of the mean radiative efficiency $\eta$.
However, within a simple disk model which neglects jet emission and other 
non-thermal phenomena, the maximum radiative efficiency is the one of the 
Novikov-Thorne model, $\eta_{\rm NT} = 1 + u_t^{\rm ms}$. For most objects, 
this should provide a conservative estimate, as the actual value is likely 
lower. So, an estimate of $\eta$ could provide a lower bound for $a_*$. 
There are several sources of uncertainty in the final result, but a mean radiative 
efficiency $\eta > 0.15$ seems to be a conservative lower limit~\cite{ref1}. For 
instance, the authors of Ref.~\cite{ref2} find a mean radiative efficiency 
$\eta \approx 0.30-0.35$ without some important assumptions necessary in 
the original version of the SoltanÕs argument.

As here we are interested in a conservative bound on the maximum value of the spin
parameter of the super-massive BH candidates, we can adopt the observational
constraint $\eta > 0.15$ from the Soltan's argument. The mean radiative efficiency
of AGNs claimed in Ref.~\cite{ref2} or the analyses of the K$\alpha$ iron line of
the BH candidates MGC-6-30-15 and 1H~0707-495 would clearly provide a
stronger bound in our study. Even if these AGNs are accreting from a thin disk, 
they may have experienced a period of super-Eddington accretion in the recent 
past, and the value of their spin parameter may be between the equilibrium 
value of a thin disk and the one of a thick disk. The allowed region on the
spin-deformation parameter plane is thus the overlap region determined by
the bounds $\eta > 0.15$ and $a_* < a_*^{\rm eq, \, thick}$, shown in Fig.~\ref{f-3}. 
The maximum value for the spin is
\be\label{eq-bo}
a_*^{\rm max} \approx 1.3 \, ,
\ee
for both JP and MMS space-times. Such a bound is only slightly weaker than 
$a_*^{\rm max} \approx 1.2$ found in Ref.~\cite{spin2}, whose discussion was 
limited to thin disks. A more detailed list of the possible bounds on $a_*$,
even assuming more stringent constraints on $\eta$, is reported in Tab.~\ref{tab}.

As already mentioned above, our argument cannot be directly applied to
stellar-mass BH candidates, as the mass gained from accretion is not relevant for
these objects and the value of their spin parameter may still reflect the natal one.
However, if we believe that stellar-mass BH candidates have been spun up by
the quick accretion of the material around them left by the explosion of the progenitor
star, the thick accretion disk model might provide even for these objects an
estimate of the maximum spin. As there are objects like GRS~1915+105, whose
radiative efficiency has been measured to be $\eta > 0.234$, the
constrain on $a_*^{\rm max}$ would be stronger than the one in~(\ref{eq-bo}),
see Tab.~\ref{tab}.

\begin{table}
\begin{center}
\begin{tabular}{c c |c c c |c c c}
\hline & & & & & & \\
Bound & \hspace{.3cm} & JP background & & & MMS background & \\
& \hspace{.3cm} & Thin disks & Thick disks & & Thin disks & Thick disks \\
\hline
$\eta > 0.15$ & & $|a_*| < 1.196 $ & $|a_*| < 1.292$ & & $|a_*| < 1.179 $ & $|a_*| < 1.312 $ \\
$\eta > 0.20$ & & $|a_*| < 1.100 $ & $|a_*| < 1.169$ & & $|a_*| < 1.090 $ & $|a_*| < 1.193 $ \\
$\eta > 0.25$ & & $|a_*| < 1.047 $ & $|a_*| < 1.092$ & & $|a_*| < 1.040 $ & $|a_*| < 1.121 $ \\
\hline
\end{tabular}
\end{center}
\caption{Constraints on the spin parameter of the super-massive BH candidates
in galactic nuclei from thin and thick accretion disks, for the case of JP and MMS
backgrounds.}
\label{tab}
\end{table}

\section{Summary and conclusions \label{s-c}}

A fundamental limit for a BH in 4-dimensional General Relativity is the bound
$|a_*| \le 1$, which is the condition for the existence of the event horizon.
However, the process of accretion can spin the object up to a spin value very
close to this limit. In the case of a thin accretion disk, the Thorne's limit is
$a_* = 0.9978$~\cite{th}. The accretion process from a thick disk may be
more efficient and spins the BH up to a spin value closer to 1. In this work, we
have considered the possibility that astrophysical BH candidates are not the
BHs predicted by General Relativity, which is not in conflict with current
observations, as the nature of these objects has still to be verified. We have
studied the accretion process from thick disks in the Polish doughnut framework
with constant angular momentum $\lambda$. The picture of the accretion
process may be qualitatively different from the one in the Kerr background,
as the marginally bound orbit may not exist. The disk may have two cusps, one
above and one below the equatorial plane, and the gas may plunge onto the
compact object from the cusps to the poles of the compact object. We have
then estimated the equilibrium value of the spin parameter of these non-Kerr
objects as a function of their deformation parameter. Finally, considering that
the radiative efficiency of AGNs should be $\eta > 0.15$, we used the argument
of Ref.~\cite{spin2} to conclude that the maximum value of the today spin
parameter of the super-massive BH candidates at the centers of galaxies is
\be
a_*^{\rm max} \approx 1.3 \, .
\ee
While we have no rigorous proof of this bound, we think it does not depend
very much on the exact background metric: a quite similar constraint was
obtained considering two different space-times, suggesting that it may 
hold regardless of the exact nature of these objects.

%%%%%%%%%%%%%%%%%%%%%%%%%%%%%%%

\begin{acknowledgments}
We thank Shiyi Xiao for useful discussions and suggestions.
This work was supported by the Thousand Young Talents Program
and Fudan University.
\end{acknowledgments}

%%%%%%%%%%%%%%%%%%%%%%%%%%%%%%

\end{document}